\documentclass{article}

\usepackage{epsfig} 
\newlength{\abstractwidth} 
\renewcommand{\thefootnote}{\fnsymbol{footnote}} 
\renewcommand{\thanks}[1]{\footnote{#1}} 
\newcommand{\starttext}{ 
\setcounter{footnote}{0} 
\renewcommand{\thefootnote}{\arabic{footnote}}} 
\renewcommand{\theequation}{\thesection.\arabic{equation}} 
\newcommand{\bea}{\begin{eqnarray}} 
\newcommand{\eea}{\end{eqnarray}} 
\newcommand{\beq}[1]{\begin{equation} \label{#1}} 
\newcommand{\be}{\begin{equation}} 
\newcommand{\ee}{\end{equation}}

\def\12{{1 \over 2}} 
\newcommand{\half}{{1\over 2}}

\newcommand{\Oline}{ \mathbf{O}   \llap{ I } }

\begin{document} 
\renewcommand{\theequation}{\thesection.\arabic{equation}} 
\bigskip
\centerline{\Large \bf {Linear Spinor Fields in Relativistic Dynamics}}
\bigskip
\begin{center} 
{\large James Lindesay\footnote{Permanent address, 
Computational Physics Lab, Department of Physics, Howard University, Washington, DC 20059}
} \\
Stanford Institute for Theoretical Physics
\end{center}
\bigskip\bigskip 
\begin{abstract} 

Linear spinor fields are a generalization of the Dirac field that have transparent
cluster decomposability properties needed for classical correspondence of
relativistic quantum systems.  The algebra of these fields directly incorporate gravitation
within a group that unifies the dynamics of the same number of additional hermitian
carriers of quantum numbers as there are gauge fields in SU(3)$\times$SU(2)$\times$U(1).  They also provide a mechanism
for the dynamic mixing of massless neutrinos using a ``transverse mass"
conjugate to the affine parameter labeling translations along its
light-like trajectory, consistent with those in the standard model.

\end{abstract} 

\starttext \baselineskip=17.63pt \setcounter{footnote}{0} 

\setcounter{equation}{0}
\section{Introduction}
\indent

The Dirac equation utilizes a matrix algebra to construct a linear
relationship between the quantum operators for energy and momentum
in the equations of motion. 
The properties of evolution dynamics described using such linear operations on quantum states are straightforward,
and have direct interpretations. 
In particular, the cluster decomposability properties necessary for classical correspondence of relativistic quantum
systems is most directly realized using linear quantum operations\cite{JLFQG}\cite{LMNP}\cite{AKLN}.

It is therefore advantageous to extend the Dirac formulation to include operators
whose matrix elements reduce to the Dirac matrices for spin $1 \over 2$
systems, but generally require that the form $\hat{\Gamma}^\mu \: \hat{P}_\mu$
be a Lorentz scalar operation.  
The finite dimensional representations
of the resulting extended Poincare group of transformations can be
constructed using the little group of operations ${\mathit{D}^{\lambda'}}_\lambda$ on the
standard state vectors, defining general transformations of the form
\beq{transformation}
\hat{U}(\underline{b}) \, \left | \psi_\lambda \: \vec{a} \right \rangle \: = \:
\sum_{\lambda'}^{} \, \left | \psi_\lambda' \: \vec{z}(\underline{b}; \vec{a}) \right \rangle
\, {\mathit{D}^{\lambda'}}_\lambda (\underline{b}; \vec{a} ) .
\ee

A spinor field equation will be demonstrated for configuration space
eigenstates of the operator $\hat{\Gamma}^\mu \: \hat{P}_\mu$ of the form
\be
\mathbf{\Gamma}^\beta \cdot {\hbar \over i} { \partial \over \partial x^\beta} \,
\hat{\mathbf{\Psi}}_{(\gamma)}^{(\Gamma)}
(\vec{x}) = -(\gamma) m c  \, \hat{\mathbf{\Psi}}_{(\gamma)}^{(\Gamma)}(\vec{x}) .
\label{LinearConfigurationSpinorFieldEqn}
\ee
For the $\Gamma=\half$ representation, the matrix representations of the
operators $\mathbf{\Gamma}^\beta$ are just one half of the Dirac matrices,
and the particle type label takes values $\gamma = \pm \half$.

\section{An Extension of the Lorentz Group} 
\indent 

The finite dimensional representations of an extension of the Lorentz group will be
constructed by developing a spinor representation of the algebra.
The group elements $\underline{b}$ will include
3 parameters representing angles, 3 boost parameters, and 4
group parameters $\vec{\omega}$ associated with four operators $\hat{\Gamma}^\mu$.

\subsection{Extended Lorentz Group Commutation Relations} 

For the extended Lorentz group, the commutation relations for angular momentum
and boost generators remain unchanged from those of the standard Lorentz group. 
The additional extended group commutation relations will be chosen to be consistent
with the Dirac matrices as follows:
\bea
\left [ \Gamma^0 \, , \, \Gamma^k \right] \: = \: i \, K_k  ,\\
\left [ \Gamma^0 \, , \, J_k \right] \: = \: 0  ,\\
\left [ \Gamma^0 \, , \, K_k \right] \: = \: -i \,  \Gamma^k  ,\\
\left [ \Gamma^j \, , \, \Gamma^k \right] \: = \: -i \, \epsilon_{j k m} \, J_m  ,\\
\left [ \Gamma^j \, , \, J_k \right] \: = \: i \, \epsilon_{j k m} \, \Gamma^m  ,\\
\left [ \Gamma^j \, , \, K_k \right] \: = \: -i \, \delta_{j k} \, \Gamma^0  .
\label{ExtLorentzGroupEqns}
\eea

A Casimir operator can be constructed for the extended Lorentz (EL) group 
in the form
\be
C \: = \: \underline{J} \cdot \underline{J} \,-\, \underline{K} \cdot \underline{K}
\,+\, \Gamma^0 \, \Gamma^0 \,-\, \underline{\Gamma} \cdot \underline{\Gamma} .
\ee
This operator can directly be verified to commute with all generators of the group. 
The operators $C$,  $\Gamma^0$, and  $J_z$ will be chosen as the set of mutually
commuting operators for the construction of the finite dimensional representations.

\subsection{Group metric of the extended Lorentz group\label{subsec:LorentzGrMetric}}

The group metric  for the algebra represented by
\be
\left [ \hat{G}_r \, , \, \hat{G}_s \right ] \: = \: -i \, \left ( c_s \right ) _r ^m \, \hat{G}_m
\ee
can be developed from the adjoint representation in terms of the structure constants: 
\be
\eta_{a b} \: \equiv \: \left ( c_a \right )_r ^s \, \left ( c_b \right )_s ^r .
\label{GroupMetricEqn}
\ee
The non-vanishing components of the extended Lorentz group metric are
given by
\bea
\eta^{(EL)} _{J_m \, J_n} \: = \: -6 \, \delta_{mn} \quad , \quad &
\eta^{(EL)} _{K_s \, K_n} \: = \: +6 \, \delta_{mn} \quad , \quad &
\eta^{(EL)} _{\Gamma^\mu \, \Gamma^\nu} \: = \: +6 \, \eta_{\mu \nu} \, .
\eea
It is important to note that the group structure of the extended Lorentz
group generates the Minkowski metric $\eta_{\mu \nu}$ as a direct
consequence of the group structure.  Neither the group structure of
the usual Lorentz group nor that of the Poincare group can \emph{generate}
the Minkowski metric as a metric describing an invariance
due to the abelian nature of the generators for
infinitesimal space-time translations.

\subsection{Symmetry Behavior of Spinor Forms}

The substitution of the
angular momenta $J_k$ and $\Gamma^0$ with barred operators
identified as
\bea
\underline{J} \: \leftrightarrow \: \underline{\bar{J}} \\
\Gamma^0 \: \leftrightarrow \: -\bar{\Gamma}^0
\eea
will preserve the commutation relations (\ref{ExtLorentzGroupEqns}).
For the Dirac case, this is seen to
represent a "particle-antiparticle" symmetry of the system, and it
represents a general symmetry under negation of the eigenvalues of
the operator $\Gamma^0$.

\subsection{Finite dimensional spinor representations} 

Spinor representations of this extension of the Lorentz group will next be constructed.

\subsubsection{Number of States}

The order of the spinor polynomial of the finite dimensional state
of the $\Gamma=J_{max}$ representation
can be determined by examining the minimal state from which other
states can be constructed using the raising operators and orthonormality.  This
minimal state takes the form
\be
\psi_{-\Gamma , -\Gamma} ^{(\Gamma)} \: = \: A^{(\Gamma J)} \chi_- ^{(-) 2 \Gamma} ,
\ee
where the $\chi_\pm ^{(\pm)}$ represent the four spinor states.  
The lower index $\pm$ labels the angular momentum basis, while the upper
index labels eigenvalues of $\Gamma^0$. 
The general state involves spinor products of the type
\be
\chi_+ ^{(+) a} \chi_- ^{(+) b} \chi_+ ^{(-) c} \chi_- ^{(-) d}.
\ee
A complete basis of states requires then that $a+b+c+d=2 \Gamma$.  By direct counting
this yields the number of states for a complete basis:
\be
N_\Gamma \: = \: {1 \over 3} (\Gamma + 1) (2 \Gamma + 1) (2 \Gamma + 3).
\ee
For instance, $N_0 = 1, N_{1 \over 2} = 4, N_1 = 10, N_{3 \over 2} = 20$, and so on.

A single J basis with $(2 J + 1)^2$ states does not cover this space of spinors.  However,
one can directly verify that
\be
N_{J_{max}} \: = \: \sum_{J=J_{min}}^{J_{max}} (2 J + 1)^2 ,
\ee
where $J_{min}$ is zero for integral systems and $1 \over 2$ for half integral systems.
Thus one can conclude that $\Gamma$ represents the maximal angular momentum state of the system:
\be
J \: \le \: \Gamma = J_{max} .
\ee
Higher order representations will include states of differing angular momenta
with common quantum statistics.

\subsubsection{Spinor metrics}

Invariant amplitudes are usually defined using dual spinors so that
the inner product is a scalar under the
transformation rule for the spinors $\mathbf{D}$:
\be
<\bar{\psi} | \phi > \: = \: <\bar{\psi'} | \phi'> \textnormal{, or  } ~
\psi_a ^\dagger g_{ab} \phi_b \: = \: 
\left( D_{ca} \psi_a  \right) ^\dagger g_{cd} 
\left( D_{db} \psi_b  \right) .
\ee
This means the \emph{spinor} metric $\mathbf{g}$ should satisfy
\be
\mathbf{g} \: = \: \mathbf{D^\dagger g D} .
\ee
The Dirac conjugate spinor $\bar{\psi}\equiv \psi^\dagger \mathbf{g}$  includes this spinor metric.

The eigenvalues of the hermitian angular momentum operators $\underline{J}$ and $\Gamma^0$
are given by real numbers
\be
\underline{J} ^\dagger=\underline{J}  \quad , \quad 
\Gamma^{0 \, \dagger}=\Gamma^0 ,
\ee
which requires that the finite dimensional representations satisfy
\be
\mathbf{g \, \Gamma^0} \, = \, \mathbf{\Gamma^0 \, g} \quad , \quad
\mathbf{g \, \underline{J} } \, = \, \mathbf{\underline{J} \, g} ~ .
\ee
The spinor metric therefore takes the general form
\be
g_{a \, a'} ^{(\Gamma J)} \: = \: (-)^{\Gamma-\gamma}
 \delta_{\gamma \gamma'} \delta_{s_z,s_z'},
\ee
using the quantum number shorthand $a:\{ \gamma, \, s_z \}$.
One can show that the general form of this spinor metric
anti-commutes with the boost generator and spatial components
of the $\underline{\Gamma}$ matrices:
\be
\mathbf{g \, \underline{\Gamma} } \, = \,- \mathbf{\underline{\Gamma} \, g} \quad , \quad
\mathbf{g \, \underline{K} } \, = \, -\mathbf{\underline{K} \, g} .
\ee
For the Dirac representation, the spinor metric takes the form of the
Dirac matrix $\mathbf{\gamma}^0$.  However, the spinor metric is
not related to $\Gamma^0$ for higher spin representations.

\subsubsection{Representation of $\Gamma={1 \over 2}$ systems}

The forms of the matrices corresponding to $\Gamma={1 \over 2}$ are
expected to have dimensionality $N_{1 \over 2}=4$, and
can be expressed in terms of the Pauli spin matrices $\sigma_j$ as shown below:
\be
\begin{array}{ll}
\mathbf{\Gamma^0} \,=\, {1 \over 2} \left( \begin{array}{cc}
\mathbf{1} & \mathbf{0} \\ \mathbf{0} & -\mathbf{1} \end{array} \right)
\,=\, {1 \over 2} \mathbf{g} \quad \quad &
\mathbf{J}_j \,=\, {1 \over 2} \left( \begin{array}{cc}
\sigma_j & \mathbf{0} \\ \mathbf{0} & \sigma_j 
\end{array} \right) \\ \\
\mathbf{\Gamma}^j \,=\, {1 \over 2} \left( \begin{array}{cc}
\mathbf{0} & \sigma_j \\ 
-\sigma_j & \mathbf{0} \end{array} \right) &
\mathbf{K}_j \,=\, -{i \over 2} \left( \begin{array}{cc}
\mathbf{0} & \sigma_j \\ 
\sigma_j & \mathbf{0} \end{array} \right) 
\end{array}
\label{4x4RepresentationEqn}
\ee
The $\Gamma^\mu$ matrices can directly be seen to be proportional
to a representation of the Dirac matrices\cite{Dirac}\cite{BjDrell} . 
A representation for $\Gamma=1$ can be found in reference \cite{JLFQG}.

\section{An Extension of the Poincare Group}
\indent

Once space-time translations are included in the group algebra, these additional commutation
relations must result in a self-consistent set of generators\cite{JLFQG}.
The extended Lorentz group structure can be minimally expanded to include
space-time translations as long as all operators continue to
satisfy the algebraic Jacobi identities,
\be
[\hat{A},[\hat{B},\hat{C}]] ~+~[\hat{C},[\hat{A},\hat{B}]]~+~
[\hat{B},[\hat{C},\hat{A}]]~=~0.
\ee
An attempt to only include the 4-momentum operators
in addition to the extended Lorentz group operators
does not produced a closed group structure, due to Jacobi relations of the
type $[\hat{P}_j , [\hat{\Gamma} ^ 0 , \hat{\Gamma} ^k] ]$.  The non-vanishing of
this commutator in the Jacobi identity implies a non-vanishing commutator
between operators $\Gamma ^\mu$ and $P_\nu$, and that this commutator must connect
to an operator which then has a commutation relation with $\Gamma ^\mu$ that yields a
4-momentum operator $P_\beta$. 
Since the 4-momentum operators self-commute,
at least one additional operator, which will be referred to as $\mathcal{M}_T$,
must be introduced.

The additional non-vanishing commutators involving Jacobi consistent operators $\hat{P}_\mu$ and
$\hat{\mathcal{M}}_T$ are given by
\bea
\left [ J_j \, , \, P_k \right] \: = \: i \hbar \, \epsilon_{j k m} \, P_m ,
\label{JPeqn} \\
\left [ K_j \, , \, P_0 \right] \: = \: -i \hbar \, P_j ,
\label{KP0eqn} \\
\left [ K_j \, , \, P_k \right] \: = \: -i \hbar \, \delta_{j k} \, P_0 ,
\label{KPeqn} \\
\left [ \Gamma^\mu \, , \, P_\nu \right] \: = \: i \, \delta_\nu ^\mu \, \mathcal{M}_T  c,
\label{GamPeqn} \\
\left [ \Gamma^\mu \, , \, \mathcal{M}_T \right] \: = \: {i \over c} \, \eta^{\mu \nu} \, P_\nu ,
\label{GamGeqn}
\eea
The first three of these relations are identical to those of the Poincare group.  The final two
relations consistently incorporate the additional operator $\hat{\mathcal{M}}_T$
needed to close the algebra.

\subsection{Invariants and group metric}

As demonstrated in section \ref{subsec:LorentzGrMetric},
given the structure constants defining the commutation relationships of the generators ,
a metric for the complete group $\eta_{s \, n}$ can be developed using (\ref{GroupMetricEqn}). 
The non-vanishing group metric elements
generated by the structure constants of this extended Poincare group are
given by
\bea
\eta^{(EP)} _{J_m \, J_n} \: = \: -8 \, \delta_{m,n} &
\eta^{(EP)} _{K_m \, K_n} \: = \: +8 \, \delta_{m,n}
\eea
\be
\eta^{(EP)} _{\Gamma^\mu \, \Gamma^\nu} \: = \: 8 \, \eta_{\mu \, \nu}
\ee
where $\eta_{\mu \, \nu}$ is the usual Minkowski metric of the Lorentz group. 
The Minkowski metric is  non-trivially generated by the extended Lorentz group algebra. 
This \emph{group theoretic} metric can be used to develop Lorentz invariants using the
operators $\Gamma^\mu$.  Since $\Gamma^\mu P_\mu$ is also Lorentz invariant,
the group transformation properties of the generators $P_\mu$, as well as their
canonically conjugate translations $x^\mu$, are direct consequences of the group properties
of the extended Poincare group.  The standard Poincare group has no non-commuting operators
that can be used to connect the group structure to the metric properties of space-time
translations.

\subsection{Unitary quantum states}

A Casimir operator for the complete group can be constructed
using the Lorentz invariants given by
\be
\mathcal{C}_m \: \equiv \: \mathcal{M}_T^2 c^2 \,-\, \eta^{\beta \nu} P_\beta P_\nu .
\label{Casimir_mu}
\ee
The label $m$ in the Casimir operator $\mathcal{C}_m$ parameterizes the eigenstates
that can be developed to construct a finite dimensional representation. 
The form of this group invariant suggests that the
hermitian operator $\mathcal{M}_T$ is a
\emph{transverse mass} parameter of the state, which can have a non-vanishing value
for massless states $\eta^{\beta \nu} P_\beta P_\nu =0$.
A set of quantum state vectors that are labeled by mutually commuting operators
are given by
\be
\begin{array}{l}
\hat{\mathcal{C}}_m \, \left | m, \Gamma,  \gamma, J, s_z \right > ~=~
m^2 c^2 \, \left | m, \Gamma,  \gamma, J, s_z \right > , \\
\hat{C}_\Gamma \,\left | m, \Gamma,  \gamma, J, s_z \right > ~=~
2 \Gamma (\Gamma + 2) \, \left | m, \Gamma,  \gamma, J, s_z \right > , \\
\hat{\Gamma}^0 \, \left | m, \Gamma,  \gamma, J, s_z \right > ~=~
\gamma \, \left | m, \Gamma,  \gamma, J, s_z \right >, \\
\hat{J}^2 \, \left | m, \Gamma,  \gamma, J, s_z \right > ~=~
J(J+1) \hbar^2  \,\left | m, \Gamma,  \gamma, J, s_z \right >, \\
\hat{J}_z \, \left | m, \Gamma,  \gamma, J, s_z \right > ~=~
s_z \hbar \, \left | m, \Gamma,  \gamma, J, s_z \right >,
\end{array}
\label{StateVectorEqn}
\ee
where $m^2$ is generally a continuous real parameter, and
all other parameters are discrete.  $\Gamma$ is an integral or
half-integral label of the representation of the extended Lorentz
group, $J$ labels the internal angular momentum representation of the state,
and $J$ has the same integral signature as $\Gamma$.

Eigenvalues of the group Casimir $\hat{\mathcal{C}}_m$ and $\hat{J}^2$
will be used to label an arbitrary standard state vector. 
An additional invariant can be constructed from the pseudo-vector
\be
\begin{array}{l}
\hat{W}_\alpha ~\equiv~ i \epsilon_{\alpha \beta \mu \nu}~
\hat{\Gamma}^\beta ~\hat{\Gamma}^\mu \, \eta^{\nu \lambda} \hat{P}_\lambda , \\
\hat{W}_0 ~=~ \underline{\hat{J}} \cdot \underline{\hat{P}} , \\
\underline{\hat{W}}~=~ \underline{\hat{K}} \times \underline{\hat{P}} ~+~
\underline{\hat{J}} \hat{P}_0 ,
\end{array}
\label{SpinOperator}
\ee
where the antisymmetric tensor $\epsilon_{\alpha \beta \mu \nu}$ is
defined by
\be
\epsilon_{\alpha \beta \mu \nu} ~\equiv ~\left \{
\begin{array}{ll}
+1 & \textnormal{for } (\alpha \beta \mu \nu) \textnormal{ an even permutation
of (0,1,2,3)}, \\
-1 & \textnormal{for } (\alpha \beta \mu \nu) \textnormal{ an odd permutation
of (0,1,2,3)}, \\
~~ 0  & \textnormal{for any two indexes equal} .
\end{array}
\right .
\ee
The Lorentz invariant $\hat{W}^2 \equiv \hat{W}_\alpha \eta^{\alpha \beta}\hat{W}_\beta$
commutes with $\hat{P}_\beta$ and $\hat{\mathcal{M}}_T$, since
$[\hat{W}_\beta, \hat{P}_\mu]=0=[\hat{W}_\beta, \hat{\mathcal{M}}_T]$. 
The covariant 4-vector $\hat{W}_\alpha$ is orthogonal to the 4-momentum operator,
$\hat{P}_\mu \eta^{\mu \nu} \hat{W}_\nu=0$,
due to the antisymmetric form defining $\hat{W}_\alpha$. 
When acting upon a massive particle state at rest (which has a time-like
4-momentum), the 4-vector $W_\alpha$
is seen to be a space-like vector whose invariant length is related to the particle's
spin times its mass.

Unitary representations of general momentum states are obtained by
boosting standard states satisfying (\ref{StateVectorEqn}).  Standard massive states
have covariant 4-momentum components $\vec{p}^{(s)}=(-m c,0,0,0)$ and vanishing
eigenvalue of transverse mass $ \hat{\mathcal{M}}_T$ labeled $m_T=0$. 
Standard massless states have covariant 4-momentum components $\vec{p}^{(s)}=(-1,0,0,1)$
with an eigenvalue of transverse mass operator given by $m_T$ that need not vanish.
General Lorentz transformations on the 4-momentum eigenstates satisfy
\be
U (\mathbf{\Lambda}^{(L)}) \, | \vec{p}, (m_T), m, J, \kappa  \rangle   ~=~
\sum_{\kappa'} | (\mathbf{\Lambda}^{(L)}\vec{p}, (m_T), m, J, \kappa'  \rangle~
Q_{\kappa' \kappa}^{(s)} (\mathbf{\Lambda}^{(L)}, \vec{p}) ,
\label{QLambdaP}
\ee
where the matrices $Q_{\kappa' \kappa}^{(s)}$ with
discrete indices $\kappa$ describe the finite dimensional, unitary transformations on the
standard momentum state of the system.  The Casimir label $m$ satisfies
$m^2c^2=m_T^2c^2 - \vec{p} \cdot \vec{p}$ for a state with
4-momentum $\vec{p}$.

\subsection{Linear Wave Equation for Single Particle States}

Eigenstates of the operator $\Gamma^\mu P_\mu$ will give linear operator dispersion relations
for energy and momenta in a spinor wave equation.  The commutators of the various group generators with
this Lorentz invariant operator are given by
\be
\left [ J_k, \Gamma^\mu P_\mu \right ] \: = \: 0
\label{JDirac}
\ee
\be
\left [ K_k, \Gamma^\mu P_\mu \right ] \: = \: 0
\label{KDirac}
\ee
\be
\left [ P_\beta, \Gamma^\mu P_\mu \right ] \: = \: 
-i \mathcal{M}_T P_\beta
\label{PDirac}
\ee
\be
\left [ \mathcal{M}_T, \Gamma^\mu P_\mu \right ] \: = \: 
-i \eta^{\beta \nu} P_\beta P_\nu
\label{MTDirac}
\ee
One should note that, from (\ref{MTDirac}), the transverse mass
only commutes with
$\Gamma^\mu P_\mu$ for massless particles. 
Similarly,  from (\ref{PDirac}) the 4-momentum operator
only commutes with
$\Gamma^\mu P_\mu$ if the transverse mass vanishes. 
Therefore, only massless states can have non-vanishing transverse mass values
$m_T \neq 0$.
The transverse mass operator $\hat{\mathcal{M}}_T$ is the generator for translations of the affine parameter labeling the
trajectory of a massless particle.

Spinor forms of the quantum state vectors and matrix representations of the operators
can be developed in the usual manner:
\be
\langle \chi_a | \hat{\Gamma}^\mu | \chi_b \rangle \equiv (\mathbf{\Gamma}^\mu )_{a b}
~,~
\mathbf{\Psi}_a^{(\Gamma)}(\vec{p}, J, \kappa) \equiv
\langle \chi_a | \vec{p}, m, J, \kappa  \rangle .
\ee
This results in a momentum-space form of the spinor equation given by
\be
\mathbf{\Gamma}^\mu \hat{P}_\mu \mathbf{\Psi}_{(\gamma)}^{(\Gamma)}(\vec{p}, J, \kappa) =
(\gamma) m c  \, \mathbf{\Psi}_{(\gamma)}^{(\Gamma)}(\vec{p}, J, \kappa) ,
\label{LinSpinMomentumSpEqn}
\ee
where $(\gamma)$ is the eigenvalue of $\hat{\Gamma}^0$ for massive particle states.
It is worth noting that this linear spinor formulation does not have negative energy
solutions.  Rather, there is a sign is associated with the particle type eigenvalue $\gamma$. 
Therefore, straightforward interpretations of the energetics of particles can be made without
introducing any filled Dirac sea of fermions to prevent transitions from positive energy
states.  There is no need to introduce any additional degrees
of freedom to stabilize the ground state once radiative coupling is included.

The configuration-space representation of (\ref{LinSpinMomentumSpEqn}) takes the form
\be
\mathbf{\Gamma}^\mu {\hbar \over i} {\partial \over \partial x^\mu} 
\mathbf{\Psi}_{(\gamma)}^{(\Gamma)}(\vec{x}) =
(\gamma) m c  \, \mathbf{\Psi}_{(\gamma)}^{(\Gamma)}(\vec{x}) .
\label{LinSpinSpaceTimeEqn}
\ee 
The spinor fields satisfy microscopic causality as long as their quantum statistics is
fermionic for $\Gamma$ half-integral, and bosonic for $\Gamma$ integral. 
The transformation properties of the fields under the improper Lorentz transformations
of parity and time reversal, as well as under charge conjugation, can be developed in
a straightforward manner\cite{JLFQG}.

A form of a Lagrangian for gravitating linear spinor fields with local gauge symmetries
will next be displayed. 
One can define spinor-valued geometric matrices of the form
$\mathbf{U}^\beta (x) \equiv \mathbf{\Gamma}^{\hat{\mu}} { \partial x^\beta
\over \partial \xi^{\hat{\mu}}  }$, where the $\xi^{\hat{\mu}}$ represent
locally flat coordinates. 
A gauge covariant Lagrangian density for a particle with Casimir label $m$ is given by
\be
\mathcal{L}_{m} = {1 \over 2 \Gamma}  {\hbar c \over i} \left [
\bar{\mathbf{\Psi}}_{(\gamma)}^{(\Gamma)} \mathbf{U}^\beta  \left ( 
 \partial_\beta  - {q \over \hbar c} A_\beta^r \, i \mathbf{G}_r
 \right ) \mathbf{\Psi}_{(\gamma)}^{(\Gamma)}  -
(\textnormal{c. c.})
\right ] + {(\gamma)  \over \Gamma} m c^2 \,  
\bar{\mathbf{\Psi}}_{(\gamma)}^{(\Gamma)} \mathbf{\Psi}_{(\gamma)}^{(\Gamma)} ,
\label{LinearSpinorGaugeFieldGravLagrangian}
\ee
where (c. c.) specifies the complex conjugate of the previous expression, and the
matrices $\mathbf{G}_r$are hermitian generators of the local gauge group of symmetries for
the linear spinor field $ \mathbf{\Psi}_{(\gamma)}^{(\Gamma)}$.  Such a Lagrangian form has
straightforward cluster decomposition properties when systems of mixed entanglements
are being described.

\subsection{Spinor Lie transformation algebra and the principle of equivalence}
 
Linear spinor fields are useful for describing
the micro-physics of gravitating systems for several reasons.  One useful property is their cluster
decomposition properties that allow straightforward combinations of systems
with arbitrary degrees of quantum entanglements at varying times. 
Another property is that the \emph{group} metric generated by the
operators $\hat{\Gamma}^{\tilde{\mu}}$ constructs the Minkowski metric.  Because
operators like $\hat{\Gamma}^{\tilde{\mu}}\hat{P}_{\tilde{\mu}}$ are
invariant under the Lorentz subgroup of transformations, the
components of the momenta likewise transform in a manner
consistent with this metric defining \emph{subgroup} invariants.  There is no analogous
Lorentz subgroup metric for the Poincare group, since there are no non-abelian
operators in that group to generate this metric. 
Since the 4-momentum operators $\hat{P}_{\tilde{\mu}}$ transform like basis vectors,
the invariance of $\hat{P}_{\tilde{\mu}} \, \eta^{\tilde{\mu} \tilde{\nu}}\, \hat{P}_{\tilde{\nu}}$
will have significance defining the \emph{space-time} metric.
The relevant group properties of the extended Poincare group will be further explored in
this section.

The general extended Poincare group transformation can be characterized by
the 15 parameters
$\mathcal{P}_X \equiv \{ \vec{\omega}, \mathbf{U},\mathbf{\Theta}, \vec{a}, \alpha   \}$
conjugate to generators $\{ \hat{\Gamma}^{\mu}, \hat{K}_{j}, \hat{J}_{k}, \hat{P}_{\nu},
\hat{\mathcal{M}}_T  \}$ representing ``Dirac boosts", Lorentz boosts, rotations,
space-time translations, and lightlike translations. 
For brevity, the 5 mutually commuting extended translations will be
labeled by a barred parameter
$\bar{a}\equiv \{ \alpha, \vec{a} \}$, while the set of 10
extended Lorentz group parameters will be underlined $\underline{a}=
\{ \mathbf{\Theta},  \mathbf{U},  \vec{\omega} \}$.  The overall group structure
will be examined for the  product
transformation of pure translations with pure extended Lorentz group transformation
defined by the convention $\hat{U}(\mathcal{P}_X) 
 \equiv \hat{X}(\bar{a}) \hat{W} (\underline{a})$.  A reversal
of this order results in a representation that is a similarity transformation on the
elements.

The individual subgroups have well-defined group operations within each subgroup, while
the overall group will have group operations based upon the convention:
\be
\begin{array}{c}
\hat{X}(\bar{b}) \, \hat{X}(\bar{a}) ~\equiv~ \hat{X}(\bar{\phi}_x (\bar{b}; \bar{a})) , \\
\hat{W}(\underline{b}) \, \hat{W}(\underline{a}) ~\equiv~ \hat{W}(\underline{\phi}_w (\underline{b}; \underline{a})) , \\
\hat{U}({\mathcal{P}_X} ') \, \hat{U}(\mathcal{P}_X) ~\equiv~ \hat{U}(\Phi ({\mathcal{P}_X} '; \mathcal{P}_X )) .
\end{array}
\ee
Using group properties, one can generally show that the translation group operation is independent of the initial
extended Lorentz group parameter $\underline{a}$,
$\bar{\Phi}_x (\bar{b}, \underline{b} ; \bar{a}) =
\bar{\phi}_x (\bar{b} ; \bar{\Phi}_x (\bar{I},\underline{b};\bar{a}))$, where
$I$ is the identity element.  Likewise,
the extended Lorentz group operation is independent of the final translation $\bar{b}$ using this convention,
$\underline{\Phi}_w (\underline{b}; \bar{a},\underline{a}) = 
\underline{\phi}_w (\underline{\Phi}_w (\underline{b}; \bar{a}, \underline{b}^{-1});
\underline{\phi}_w (\underline{b};\underline{a}))$.
The inverse group element to $\mathcal{P}_X=\{ \underline{a}, \bar{a} \}$, which
can be calculated using
$\hat{W}^{-1} \hat{X}^{-1} = \hat{W}^{-1} \hat{X}^{-1} \hat{W} \hat{W}^{-1}$, is given by
$\mathcal{P}_X^{-1}= \{ \underline{\Phi}_w (\underline{a}^{-1}; \bar{a}^{-1}, \underline{I}) ,
\bar{\Phi}_x (\bar{I}, \underline{a}^{-1}; \bar{a}^{-1}) \}$.

Group associativity for operations within the subgroups is expressed in the relationships
\bea
\bar{\Phi}_x (\bar{c},\underline{c} ; \bar{\Phi}_x (\bar{b},\underline{b};\bar{a}))=
\bar{\Phi}_x (\bar{\Phi}_x(\bar{c},\underline{c};\bar{b}), 
\underline{\Phi}_w (\underline{c}; \bar{b},\underline{b}); \bar{a})  \quad , 
\label{xTranslAssocConditionEqn}\\
\underline{\Phi}_w (\underline{c}; \bar{\Phi}_x (\bar{b}, \underline{b} ; \bar{a}) ,
\underline{\Phi}_w (\underline{b}; \bar{a}, \underline{a}) ) =
\underline{\Phi}_w (\underline{\Phi}_w (\underline{c}; \bar{b}, \underline{b});
\bar{a},\underline{a}) ~.
\label{xLorentzAssocConditionEqn}
\eea
In particular,
the translationally independent group of transformations
\begin{displaymath}
\bar{\Phi}_x (\bar{I},\underline{c} ; \bar{\Phi}_x (\bar{I},\underline{b};\bar{x}))=
\bar{\Phi}_x (\bar{I} , 
\underline{\Phi}_w (\underline{c}; \bar{I},\underline{b}); \bar{x}) 
\end{displaymath}
forms a Lie transformation group
$\bar{x}' \equiv \bar{f}(\bar{x};\underline{b})=
\bar{\Phi}_x (\bar{I}, \underline{b}; \bar{x})$. 
However, one should note that the full group of transformations is generally \emph{beyond}
those of a traditional Lie transformation group. 
The transformations in gauge symmetries are typically represented by
traditional Lie transformation groups.

Given the group operation, the complete set of group parameters (like
structure constants, Lie structure matrices, transformation matrices, etc.)
can be constructed. 
The matrices that define how the group transformations mix the generators of the
group $U(\mathcal{P}^{-1}) \, G_r \, U(\mathcal{P}) = 
\oplus_r{}^s (\mathcal{P}) \, G_s$ are given by\cite{JLFQG}
\be
\oplus_r{}^s (\mathcal{P}) = \left . {\partial \over \partial \mathcal{P}^r{} '}
\Phi^s (\mathcal{P}^{-1} ; \Phi (\mathcal{P}' ; \mathcal{P}))
\right |_{\mathcal{P}' \rightarrow I} =
\left . {\partial \Phi^m (\mathcal{P}' ; \mathcal{P})) \over \partial \mathcal{P}^r{}'} \right |_{\mathcal{P}' =I}
\left . {\partial \Phi^s (\mathcal{P}^{-1} ; \mathcal{P}')) \over \partial \mathcal{P}^m{} '}
\right |_{\mathcal{P}' =\mathcal{P}},
\ee
where it is convenient to define the Lie transformation matrices
$\Oline_{m}{}^{s} (\mathcal{P}) \equiv
\left . {\partial \Phi^s (\mathcal{P}^{-1} ; \mathcal{P}')) \over \partial \mathcal{P}^m{} '}
\right |_{\mathcal{P}' =\mathcal{P}}$,
and $\Theta_r{}^m (\mathcal{P}) \equiv
\left . {\partial \Phi^m (\mathcal{P}' ; \mathcal{P})) \over \partial \mathcal{P}^r{}'} \right |_{\mathcal{P}' =I}$. 
With these definitions, the  transformation matrices for the generators satisfy
$\oplus_r{}^s (\mathcal{P})=\Theta_r{}^m (\mathcal{P}) \Oline_{m}{}^{s} (\mathcal{P}) $.

The transformations $\bar{\phi}_x (\bar{b}; \bar{a})=\bar{\Phi}_x (\bar{b}, \underline{I}; \bar{a})$
form an abelian subgroup of translations.
The extended translations all commute with each other, but they are mixed amongst each other
by the extended Lorentz transformations.  This means that a general operation of the form
$F(\xi^\Lambda \hat{P}_\Lambda)$ will transform under extended Lorentz transformations
according to
\be
\hat{U}(\{\underline{a},\bar{I}\}) \, F(\xi^\Lambda \hat{P}_\Lambda) \,
\hat{U}^{-1} (\{\underline{a},\bar{I}\}) =
F(\xi^\Lambda \, \oplus_\Lambda{}^\Delta (\underline{a}^{-1}) \hat{P}_\Delta) ,
\ee 
where the capital Greek indices sum over the five parameters including the space-time coordinates and the affine
coordinate conjugate to the transverse mass. 
Thus, the $\oplus_\Lambda{}^\Delta (\underline{a})$ define the extended Lorentz transformation
matrices on the momenta. 
The fact that the translations are abelian allows a very useful choice of the
translation group parameters associated with space-time coordinates. 
If one utilizes the factor 
$\Oline_{\Delta}{}^{\Lambda} (\bar{x}) \equiv {\partial  \over
\partial x^\Delta} \phi_x^{\Lambda} (\bar{x}^{-1};\bar{x})$,
the associativity condition (\ref{xTranslAssocConditionEqn})
implies that
\bea
\Oline_{\Delta}{}^{\Lambda} (\bar{x})  \oplus_\Lambda{}^\Upsilon (\bar{a})=
{\partial \phi_x^\Lambda (\bar{x}; \bar{a}) \over \partial x^\Delta}
\Oline_{\Lambda}{}^{\Upsilon} (\bar{\phi}_x  (\bar{x}; \bar{a})) \Rightarrow \qquad \nonumber \\ 
dx^\Delta \Oline_{\Delta}{}^{\Lambda} (\bar{x})  \oplus_\Lambda{}^\Upsilon (\bar{a}) =
d \phi_x^\Lambda (\bar{x}; \bar{a}) \Oline_{\Lambda}{}^{\Upsilon} (\bar{\phi}_x(\bar{x}; \bar{a}) ) \, .
\eea 
Therefore, if one defines the special set of coordinates $\xi^{\tilde{\Upsilon}}$ by
\be
{\partial \xi^{\tilde{\Upsilon}} \over \partial x^\Delta} \equiv \Oline_{\Delta}{}^{\tilde{\Upsilon}} (\bar{x})  ~,
\ee
then these coordinates have the property that
\bea
\xi^{\tilde{\Upsilon}} (\bar{a}) = \int_{\bar{I}}^{\bar{a}} dx^\Delta \Oline_{\Delta}{}^{\Upsilon} (\bar{x})
~ , \qquad \qquad \\
\xi^{\tilde{\Delta}} (\bar{b})  \oplus_{\tilde{\Delta}}{}^{\tilde{\Upsilon}} (\bar{a}) =
 \int_{\bar{a}}^{\bar{\phi}_x (\bar{b};\bar{a})} d \phi_x^\Delta \, \Oline_{\Delta}{}^{\tilde{\Upsilon}} (\bar{\phi}_x).
\eea
This means that these coordinates satisfy
$\xi^{\tilde{\Upsilon}} (\bar{\phi}_x (\bar{b};\bar{a}))=
\xi^{\tilde{\Delta}} (\bar{b})  \oplus_{\tilde{\Delta}}{}^{\tilde{\Upsilon}} (\bar{a}) +\xi^{\tilde{\Upsilon}} (\bar{a}) $.
The coordinate transformation is related to the group operation via
\be
{\partial \xi^{\tilde{\Upsilon}}(\bar{x}) \over \partial x^\Delta} =\mathcal{V}^{\tilde{\Upsilon}}{}_\Delta (\bar{x}) =
\left . {\partial \Phi_x^{\tilde{\Upsilon}} (\bar{x}^{-1},\underline{I};\bar{x}') \over \partial x'{} ^\Delta}
\right |_{\bar{x}'=\bar{x}} \, .
\ee
This equation directly relates the tetrads $\mathcal{V}^{\tilde{\mu}}{}_\beta$
to the extended Poincare group operation.

More generally, the special coordinates satisfy
\be
\xi^{\tilde{\Upsilon}} ( \bar{\phi}_x (\bar{x}_2, \underline{a}_2 ; \bar{x}_1)) = 
\xi^{\tilde{\Upsilon}} (\bar{x}_2) + \xi^{\tilde{\Lambda}} (\bar{x}_1) \, 
\oplus_{\tilde{\Lambda}}{}^{\tilde{\Upsilon}} (\underline{a}_2^{-1}) \, ,
\ee
or, in a more suggestive form
\be
 \bar{\phi}_x (\bar{x}_2, \underline{a}_2 ; \bar{x}_1) = 
\bar{x}_3 (\overline{\xi (\bar{x}_2)} + \overline{\xi (\bar{x}_1)  \oplus (\underline{a}_2^{-1})} ) \, .
\label{CurvilinearFromGroupEqn}
\ee
The expression (\ref{CurvilinearFromGroupEqn}) demonstrates a direct mapping of
locally flat coordinates into curvilinear coordinates, consistent with the principle of equivalence.

\section{Dynamic mixing of massless states}
\indent

The transverse mass operator $\hat{\mathcal{M}}_T$
of a massless particle is the generator for affine parameter translations $\Delta \lambda$
along the particle's light cone trajectory, and its non-vanishing eigenvalue propagates a stationary particle 
with the usual quantum phase $e^{-{i \over \hbar} m_j c \, \Delta \lambda}$.  Since all
massless particles share the same phase for space-time propagation $e^{{i \over \hbar} \vec{p} \cdot \vec{x}}$,
the usual manner for differentiating particles for dynamic mixing requires the introduction of
small masses for the particles\cite{PDG}.  However, massless particles of differing transverse mass can
be mixed in a straightforward manner.  In particular, a mechanism for mixing massless neutrinos of fixed helicity
$\pm {1 \over 2} \hbar$ will be developed.

Suppose that massless neutrinos of finite transverse mass mix to form
flavor eigenstates in a manner consistent with the single helicity states
giving V-A couplings in weak interactions, and
analogous to the quark mixing that suppresses
neutral, strangeness-changing currents.  The transverse mass eigenstates
will be labeled by $| m_j \rangle$, while the eigenstates of flavor that define generation
$a$ will be labeled $| f_a \rangle$.  Any mixing due to the dynamics is expected
to relate the states in a manner that preserves unitarity,
\be
| f_a \rangle ~=~ \sum_{j} | m_j \rangle \, U_{j \, a} ,
\ee
requiring that the components $U_{j \, a}$ define a unitary matrix.
Finite translations for massless
particles take the form of a simple exponential
in terms of the affine parameter along the trajectory $\lambda$, and the generator of infinitesimal
affine translations $\hat{\mathcal{M}}_T$, i.e. $T_\lambda = e^{-{i \over \hbar} \lambda  \, \hat{\mathcal{M}}_T c}$.
This means that the transition amplitude for mixing massless flavor eigenstates
$f_a \rightarrow f_b$ is of the form
\be
A(f_b \leftarrow f_a) ~=~ \sum_ j   U_{j \, b}^* ~
e^{-{i \over \hbar} m_j c \, \Delta \lambda} ~  U_{j \, a}. 
\ee
The scale of the affine parameter is given by the spatial/temporal distance
of the null particle trajectory $\Delta \lambda = L=c T$.

The transition probability for the mixing $\mathcal{P}(f_b \leftarrow f_a)=
|A(f_b \leftarrow f_a)|^2$ satisfies
\bea
\mathcal{P}(f_b \leftarrow f_a)~=~\delta_{b \, a} \quad + \qquad \qquad  \qquad \qquad  \qquad \qquad 
  \qquad \qquad \qquad \qquad  \qquad \qquad \nonumber  \\
-2 \sum_{j<k}\left \{ 2 Re[ \Upsilon_{j \, k}(b , a)] sin^2 \left ( {\delta m_{j  k}\, c\, L \over 2 \hbar} \right ) +
Im[ \Upsilon_{j \, k}(b , a)] sin \left ( {\delta m_{j  k}\, c\, L \over  \hbar} \right )
\right \} , ~
\eea
where $\Upsilon_{j \, k}( b, a) \equiv U_{j \, b} \, U_{j \, a}^* \, U_{k \, b}^* \, U_{k \, a}$.
Thus, massless particles with differing transverse
mass eigenvalues $\delta m_{j  k}=m_j - m_k$ can indeed allow dynamical mixing of flavor eigenstates.

\section{Additional Hermitian generators}
\indent

The fundamental representation of the extended Lorentz group can be developed in terms
of $4 \times 4$ matrices, with a particular representation given in
(\ref{4x4RepresentationEqn}).  The three angular momentum generators, along with $\Gamma^0$,
make up the 4 Hermitian generators of this group.  It is of interest to examine the other Hermitian
generators in the group GL(4).

There are 16 Hermitian generators whose representations are  $4 \times 4$ matrices.  This
means that there are 12 additional $4 \times 4$ Hermitian generators beyond those of
the extended Lorentz group.  One of these generators is proportional to the identity
matrix, and thus commutes with all other generators.  Thus, this generator forms
a U(1) internal symmetry group that defines a conserved hypercharge on the algebra.

Three additional generators $\tau_j$ form a closed representation of SU(2) on the lower components
of a spinor:
\be
\tau_j ={1 \over 2} \left (
\begin{array}{cc}
\mathbf{0} & \mathbf{0} \\
\mathbf{0} & \mathbf{\sigma}_j
\end{array}
\right ) .
\ee
These generators transform as components of a 3-vector under the little group of transformations for
a massive particle.  
In a Lagrangian of the form (\ref{LinearSpinorGaugeFieldGravLagrangian}), transformations involving
these generators vanish on upper component standard state vectors.

Six additional generators are given by the Hermitian forms of the anti-Hermitian generators
$\mathbf{\Gamma}^j$ and $\mathbf{K}_j$ given by
$\mathbf{T}_j = i \,  \mathbf{\Gamma}^j$ and 
$\mathbf{T}_{j+3} = i \, \mathbf{K}_j$.  The final two generators are given by
\be
\mathbf{T}_{7} ={i \over 2} \left (
\begin{array}{cc}
\mathbf{0} & \mathbf{1} \\
-\mathbf{1} & \mathbf{0}
\end{array}
\right )
\quad , \quad
\mathbf{T}_{8} ={1 \over 2} \left (
\begin{array}{cc}
\mathbf{0} & \mathbf{1} \\
\mathbf{1} & \mathbf{0}
\end{array}
\right )  \, .
\ee
The set of 8 Hermitian generators $\mathbf{T}_s$ do not form a closed algebra independent of the other Hermitian
generators.  The spinor metric transforms these generators according to
$\mathbf{g}\mathbf{T}_s \mathbf{g}=-\mathbf{T}_s$. 
Transformations involving at most 5 re-combinations of these generators
will vanish on upper component standard state vectors for a Lagrangian
of the form  (\ref{LinearSpinorGaugeFieldGravLagrangian}).  However, the remaining three combinations
necessarily mix particle states defined by the $\Gamma^\mu P_\mu$ form of the Lagrangian.

\section{Conclusion}
\indent

Physical models that are unitary, maintain quantum linearity, have positive definite energies, and have
straightforward cluster decomposition properties, can be constructed in
a straightforward manner using linear spinor fields. 
The piece of the group algebra that connects the group structure to metric gravitation
necessitates the inclusion of an additional group operator that generates affine parameter
translations for massless particles.  This allows dynamic mixing of massless particles
in a manner not allowed by the standard formulations of Dirac or Majorana.

Just as classical mechanics emerges from the expectation values of quantum processes,
space-time geometry can be assumed to emerge from expectation valued measurements of quantum energies
and momenta via Einstein's equation.  Using this interpretation, classical geometrodynamics is emergent from
the behaviors of ensembles of mixed quantum states as they independently decohere. 
Linear spinor fields then maintain their linearity by describing coherence using 
coordinate descriptions transformed from the proper coordinates of the gravitating fields.

Conserved particle type can be shown to be a consequence of the spinor field equation. 
Internal gauge symmetries can be incorporated into the formulation in the usual manner. 
In addition,
the fundamental representation of the linear spinor fields unifies a set of micro-physical interactions
involving quanta exchanging 12 Hermitian degrees of freedom, with the geometrodynamics
of general relativity through a single unified group of transformations.  At least some of these interactions
necessarily mix fundamental standard states of representations.  It is intriguing that the
the number of additional Hermitian degrees of freedom beyond those defining the extended Lorentz
group is the same as the number of generators in the standard model of fundamental interactions.

\section{Acknowledgements}

The author wishes to acknowledge the support of Elnora Herod and
Penelope Brown during the intermediate periods prior to and after
his Peace Corps service (1984-1988), during which time the bulk of this work was
accomplished.  In addition, the author wishes to recognize the
hospitality of the Department of Physics at the University of Dar
Es Salaam during the three years from 1985-1987 in which a substantial portion of
this work was done.  Finally, the author wishes to express his appreciation of the
hospitality of the Stanford Institute for Theoretical Physics during his year of
sabbatical leave.

\end{document}